\documentstyle[floats,twocolumn,prl,aps]{revtex} 

\input epsf 
\input rotate 
\begin{document} 
 
\title{ 
Phase-Field Model of Dendritic Sidebranching with Thermal Noise} 
 
\author{Alain Karma$^{(a)}$ and Wouter-Jan Rappel$^{(b)}$ } 
 
\address{$^{(a)}$Department of Physics and Center  
for Interdisciplinary Research 
on Complex Systems,\\ Northeastern University, 
Boston, MA 02115\\ 
$^{(b)}$Department of Physics, University of California, 
San Diego, La Jolla, CA 92093} 

\date{January 29, 1999}
 
\author{\parbox{397pt}{\vglue 0.3cm \small 
We investigate dendritic sidebranching during crystal growth 
in an undercooled melt by simulation of 
a phase-field model which incorporates 
thermal noise of microscopic origin. 
As a non-trivial quantitative test 
of this model, we first show that the simulated
fluctuation spectrum of a 
one-dimensional interface in thermal equilibrium agrees
with the exact sharp-interface spectrum 
up to an irrelevant short-wavelength cutoff comparable 
to the interface thickness. 
Simulations of dendritic growth  
are then carried out in two dimensions to compute 
sidebranching characteristics (root-mean-square amplitude 
and mean spacing between sidebranches) as a function of  
distance behind the tip. These quantities are found 
to be in good overall quantitative agreement with the 
predictions of the existing linear WKB theory of noise amplification. 
The extension of this study to three dimensions  
remains needed to determine 
the origin of noise in experiments. 
}} 
\maketitle 
 
\section{INTRODUCTION} 
 
Dendrites are intricate
growth patterns that  
make up the microstructure of  
many important commercial alloys \cite{KurFis,KurTri:DenRev}.
They develop a complex shape
due to the emission of secondary branches behind the  
growing tips of primary branches \cite{HG}.  
A major advance in  
understanding this dynamical process 
came historically from the insight \cite{Pel,PelCla}
that results of Zel'dovich {\it et al.} \cite{Zeletal}
on the stability of flame fronts could be extended
to other interfacial pattern forming systems  
such as dendrites, viscous fingers, etc.
For dendrites, further developments along
this line \cite{Pie,BBL,KL1,Lan,Bre:noi}
led to a physical picture where 
small noisy perturbations,  
localized initially at the tip,  
become amplified to a macroscale along the sides  
of steady-state needle crystals, thereby 
giving birth to sidebranches \cite{Pie,BBL,KL1,Lan,Bre:noi}
in qualitative agreement with some
experiments \cite{DKP}.
 
This sidebranching mechanism 
requires some continuous source of
noise at the tip.
Therefore, thermal noise, originating
from microscopic scale fluctuations 
inherent in bulk matter, is the most
natural and quantifiable candidate to consider.
Langer \cite{Lan} analyzed 
the amplification of thermal noise along the  
sides of an axisymmetric paraboloid of revolution  
and concluded from a rough estimate that it 
is probably not strong enough to 
explain experimental observations, i.e. sidebranches form 
closer to the tip in experiment than predicted on the basis 
of thermal noise amplification. More recently, Brener and 
Temkin \cite{Bre:noi} made the interesting observation that noise 
is amplified faster along the more gently sloping 
sides of anisotropic (non-axisymmetric) needle crystals, 
leading to the conclusion that thermal noise has  
about the right magnitude to fit experimental data. 
 
There remain, however, several sources of uncertainties 
regarding this conclusion. Firstly, calculations of 
noise amplification have been based on a WKB 
(Wentzel-Kramers-Brillouin) approximation 
which has only been tested by
comparison \cite{BBL} with numerical simulations \cite{KL1}
for a fixed frequency 
perturbation localized at the tip.
Thermal noise is more 
difficult to analyze because it involves a 
wide range of frequencies and is
spatially distributed.
Consequently, current estimates of the sidebranching 
amplitude \cite{Lan,Bre:noi}
involve some overall prefactor which is only known approximately. 
Secondly, the predicted sidebranching amplitude 
depends sensitively on the non-axisymmetric tip shape
which seems to vary from system to system. 
Bisang and Bilgram \cite{BisBil}
have found that the tip of Xenon dendrites is well fitted
by the power law $x\sim z^{3/5}$ \cite{Almetal,Bre:tail}
(as opposed to $x\sim z^{1/2}$ for
a paraboloid), where $z$ is the distance
behind the tip and $x$ is the radial distance from
the growth axis to the interface. 
In contrast, LaCombe et al. \cite{LaCetal}, 
find that the tip shape of succinonitrile dendrites is well described  
up to $10\,\rho$ (where
$\rho$ is the tip radius) behind the tip by a weak
four-fold deviation from a paraboloid 
$x\sim z^{1/2}$.
Since the $z^{3/5}$ power law should only strictly hold far
behind the tip \cite{Almetal,Bre:tail}, 
the proposal \cite{Bre:noi} that it can be used to 
predict the sidebranching amplitude remains to be validated
beyond the experiments of Bisang and Bilgram \cite{BisBil}.
Lastly, analyses of sidebranching have so far been 
constrained to a linear regime. Therefore, there 
remains the possibility that nonlinearities produce 
a noisy limit cycle where sidebranches drive 
tip oscillations.  
 
At present, it appears to be  difficult to 
make further progress 
on these issues without some reliable computational 
approach to accurately simulate  
dendritic growth with thermal noise.
Numerical simulations of dendritic  
growth using a phase-field approach 
are consistent with 
a noise amplification scenario in that sidebranches
are absent in purely deterministic simulations where
the diffuse interface region is well resolved 
\cite{Wheetal:Sim,WarBoe,WanSek:Sim,Muretal:Sim,KarRap,Proetal}. 
Moreover, in certain simulations 
sidebranching has been induced by randomly driving 
the tip \cite{Wheetal:Sim,WarBoe} in a fashion which
is adequate to produce dendritic 
microstructures, but not to investigate quantitatively the 
physical origin of sidebranching. 
In addition, in front-tracking simulations \cite{IhlMul}, sidebranching
appears to be due to the  
amplification of numerical noise which is 
difficult to control.
 
The first goal of this  
paper is to demonstrate that  
the phase-field approach 
\cite{Lan:Ma,ColLev} can be
successfully extended to study the 
effect of thermal noise quantitatively. The 
second goal is to use this approach to carry out a 
quantitative study of sidebranching in order 
to test the predictions of the linear WKB theory 
of noise amplification \cite{Lan,Bre:noi}. 
Here, simulations are restricted to two dimensions 
in order to carry out this comparison in  
the simplest non-trivial test case. 
There are two 
main reasons to elect a phase-field 
approach to study thermal noise. 
Firstly, this approach has proven 
extremely successful to simulate 
dendritic growth
\cite{Wheetal:Sim,WarBoe,WanSek:Sim,Muretal:Sim,KarRap}. 
By reformulating the 
asymptotic analysis of the phase-field 
model, it has recently been possible to lower 
the accessible range of undercooling 
as well as to choose an 
arbitrary small interface kinetic coefficient \cite{KarRap}. 
In addition, adaptive mesh refinement methods, 
used in combination with the reformulated asymptotics, 
have pushed the limit of undercooling even 
further towards the experimental range \cite{Proetal}. 
Secondly, the phase-field approach provides 
a natural framework to incorporate thermal 
noise since it is adapted from phenomenological continuum models 
of second order phase transitions used to study
fluctuations near a critical point \cite{Haletal}. 
Therefore, the formalism to incorporate 
noise into such models already exists.  
The extension to the phase-field model 
mainly requires the use of the  
fluctuation-dissipation theorem together with 
an appropriate scaling of parameters 
to relate the magnitude of the noise  
in the model with the noise 
that is present in an experiment. This 
straightforward exercise is carried out here.
An additional issue is the 
numerical resolution of a small amplitude noise 
which could be masked  
by the numerical noise and/or 
discretization artifacts that  
are present in simulations. This problem  
is absent in studies of phase-transitions  
where the bare magnitude of the noise 
is not important. Here, however, this magnitude plays a crucial role. 
Fortunately, we shall find that it is possible to  
resolve accurately a small amplitude noise, of 
magnitude comparable to experiment, provided that the 
spatially diffuse interface region is  
well resolved. 
 
In the context of this study, we are 
naturally led to revisit the issue of  
the relative importance of the 
noises acting in the bulk 
and at the interface, which was 
previously considered in the context of a  
sharp-interface model \cite{Karma}. 
Microscopically, the 
bulk noise originates from fluctuations 
in the heat current in the solid and 
liquid phases, whereas the interface noise  
originates from the exchange  
of atoms between the two phases 
(i.e. the attachment and detachment of 
atoms at the interface).  
In ref. \cite{Karma},  
it was shown by a direct calculation of the 
equilibrium fluctuation spectrum of a flat interface 
that the bulk and interface noises drive, respectively, 
long-wavelength ($\lambda>\lambda^*$)  
and short-wavelength ($\lambda<\lambda^*$) regions of 
this spectrum, where the cross-over length  
$\lambda^*=4\pi c D/\mu L$. Here, $c$ is the specific 
heat per unit volume, $D$ the thermal diffusivity, $L$ the latent 
heat of melting per unit volume, and $\mu$ the interface kinetic 
coefficient. On this basis, it was roughly estimated that the 
bulk noise should predominantly drive sidebranching whenever  
$\lambda^*<\lambda_S$, where, $\lambda_S\sim \sqrt{Dd_0/V}$, 
is the stability length below which perturbations of the interface 
are stable, $V$ is the tip velocity, and $d_0$ is the capillary length. 
This condition is actually satisfied for growth at 
low velocity where simple estimations allow 
one to conclude that $\lambda_S\gg \lambda^*$ for 
materials with reasonably fast attachment kinetics. 
In the phase-field model, the  
bulk and interface noises are represented by 
Langevin forces added to 
the evolution equations for the temperature  
and phase fields, respectively. It it therefore 
possible to probe the relative importance 
of these two noises. In this paper, we 
focus on a low velocity limit
where the bulk noise should 
be dominant according to the above estimate.  
We observe indeed hat sidebranching is unaffected 
when the noise is switched off in the evolution equation for 
the phase-field, and only the conserved  
noise is kept in the diffusion equation.
 
This paper is organized as follows.  
In Section II, we review the sharp-interface  
equations of solidification with thermal noise and 
certain useful results of fluctuation theory. 
In section III, we introduce the phase-field model 
and analyze its equilibrium fluctuation properties, which allows  
us to relate the parameters of this model to 
the known material parameters that enter in the 
sharp-interface model. In section IV we then 
discuss the numerical implementation of 
the model and present the results of 
a detailed numerical test based on comparing 
the simulated and analytically predicted fluctuation spectra 
of a stationary interface in thermal equilibrium. 
Next, in section V, we present the results of the 
simulations of dendritic growth and a quantitative comparison 
of the sidebranching characteristics (amplitude and sidebranch spacing)  
of a steady-state growing dendrite to the  
analytical predictions of the WKB theory.
Finally, concluding remarks are presented in section VI. 
 
\section{Sharp-Interface Model} 
 
We consider the standard 
symmetric model with equal thermal diffusivities in 
the solid and liquid phases. The incorporation of 
fluctuations in this model, with reference 
to earlier works, is discussed in detail in Ref. \cite{Karma} 
and we only review here the main results. 
The basic equations of the  
model are given by: 
\begin{eqnarray} 
\partial_t T& =& D\, \nabla^2 T 
-\vec \nabla \cdot {\vec j} 
\label{e1}\\ 
LV_n & =& -c\,D\, \hat n \cdot \left( \vec \nabla T|_l -\vec \nabla T|_s 
\right) 
\,+\,c\,\hat n \cdot \left(\vec j|_l- \vec j|_s \right)\label{e2} \\ 
T_I & =& T_M-\Gamma\,\kappa\, 
-\,\frac{V_n}{\mu}~+~\eta 
\label{e3} 
\end{eqnarray} 
where $T(\vec r,t)$ is the temperature field 
defined in terms of the three-dimensional 
position vector $\vec r=x\hat x+y\hat y+z\hat z$, 
$T_I$ is the interface temperature,
$T_M$ is the melting temperature,  
$\Gamma=\gamma T_M/L$ is the Gibbs-Thomson 
coefficient where $\gamma$ is the surface energy,  
$V_n$ is the normal velocity of the interface,  
$\vec \nabla T|_l$ ($\vec \nabla T|_s$) is the
temperature gradient evaluated on
the liquid (solid) side of the interface,
$\kappa$ is the interface curvature, and
other parameters were defined in section I.
The conserved noise, 
$\vec j = j_x \hat x + j_y \hat y + j_z \hat z$, 
represents the fluctuating part of the heat current, 
where the components, $j_m$, with ($m=x,y,z$), are 
random variables uncorrelated in space and time that 
obey a Gaussian distribution. The variance of this 
distribution 
\begin{eqnarray} 
& &<j_m(\vec r,t)j_n(\vec r',t')> = 
2\,\frac{Dk_BT(\vec r, t)^2}{c} \nonumber\\  
& &~~~~~~~~~~~~~~~~~~~~~~~~~~~~~~~~~ 
\times~\delta_{mn}\delta(\vec r-\vec r')\delta(t-t'),\label{e4} 
\end{eqnarray} 
is fixed by the requirement that the 
diffusion equation driven by  
this noise produces, in equilibrium, 
the known distribution of temperature fluctuations 
in the solid and liquid phases, which is a  
simple application of the fluctuation-dissipation 
theorem. According to basic 
principles of statistical physics \cite{LanLif}, 
the mean square fluctuation of  
the temperature in a small 
volume $\Delta V$ of solid or liquid  
is given by \cite{LanLif}: 
\begin{equation} 
<\Delta T^2>=\frac{k_BT_M^2}{c\Delta V}\label{tempfluc} 
\end{equation} 
where $k_B$ is the Boltzmann constant, which is precisely 
the result that one obtains from a simple calculation 
of $<\Delta T^2>$ using Eq. \ref{e1} with $\vec j$ 
defined by Eq. \ref{e4}. Note that, in a non-equilibrium 
situation, the temperature variation in the liquid is 
small compared to the melting temperature, such that 
$T(\vec r,t)$ can be replaced by $T_M$  
on the r.h.s. of Eq. \ref{e4}.  
 
Next, to write down the  
correlation of the non-conserved noise 
that enters in the interface condition (\ref{e3}), 
it is convenient to define the interface position, 
$\zeta(\vec r_\perp,t)\equiv z_{int}(\vec r_\perp,t)$, 
where $\vec r_\perp=x{\hat x}+y{\hat y}$ is the two-dimensional position 
vector in the plane perpendicular to the $z$-axis.  
The interface temperature is then simply 
given by, $T_I=T(\vec r_{int},t)$,  
where $\vec r_{int}=\vec r_\perp+\zeta(\vec r_\perp,t)\hat z$. 
We can assume, without loss of generality, that the 
interface is locally single valued (i.e. no overhang)  
with respect to this set of coordinates; 
$\eta(\vec r_\perp,t)$ is then Gaussianly distributed with  
a variance defined by 
\begin{equation} 
<\eta(\vec r_\perp,t)\,\eta(\vec r_\perp',t')>~=~ 
2\,\frac{k_B\,T_I^2}{\mu\,L} 
\frac{~\delta(\vec r_\perp-\vec r_\perp')\, 
\delta(t-t')}{\sqrt{1\,+\,|\vec\nabla_\perp \zeta(\vec r_\perp,t)|^2}} 
\label{e5} 
\end{equation} 
where the square-root in  
the denominator of Eqn. \ref{e5} is a simple  
geometrical factor introduced such that the net force on a small 
area $dS$ of the interface is independent  
of its local orientation \cite{Karma},
and $\vec\nabla_{\bot}=\hat x\partial_x 
+\hat y \partial_y$ denotes the two-dimensional  
gradient vector in the plane of the interface. 
The application of the  
fluctuation-dissipation theorem for this noise 
requires that its variance be chosen such that 
the sharp-interface model reproduces 
the known fluctuation spectrum of a stationary interface 
in thermal equilibrium, derived  
analytically in the next section (see also \cite{Karma}): 
\begin{equation} 
S(k)\equiv<\zeta_k\zeta_{-k}>=\frac{k_BT_M}{\gamma\,k^2}\label{intfluc} 
\end{equation} 
where $\zeta_k$ is the Fourier coefficient of 
the interface displacement, i.e. 
\begin{equation} 
\zeta(\vec r_\perp)=\int\,\frac{d^2k}{(2\pi)^2} 
\,e^{i\vec  k\cdot\vec r_\perp}\,\zeta_k 
\end{equation} 
A straightforward but lengthy  
calculation described in Ref. \cite{Karma} shows 
that Eqs. \ref{e1}-\ref{e3}, with the noises 
defined by Eqs. \ref{e4} and \ref{e5}, 
yields this spectrum in equilibrium. 
 
\section{Phase-field model} 
 
The Langevin formalism  
to incorporate fluctuations into continuum 
models of phase transitions is well-established \cite{Haletal}, 
and the same procedure can be followed for the 
phase-field model. Like in the sharp-interface 
model \cite{Karma}, we proceed by adding stochastic forces 
whose magnitudes are determined  
by making contact with equilibrium properties. 
For this purpose, 
it is convenient to express the phase-field 
model in terms of the dimensionless  
temperature field 
\begin{equation} 
u=\frac{T-T_M}{L/c}, 
\end{equation} 
and the local enthalpy per unit volume 
defined by, 
\begin{equation} 
H=e_0\left(u-\frac{p(\phi)}{2}\right), 
\label{Hdef} 
\end{equation} 
where $e_0$ is a constant  
with units of energy per unit volume, 
$\phi$ is the phase-field chosen to vary 
between $-1$ in the liquid and $+1$ in the solid, 
and $p(\phi)$ is some monotonously 
increasing function of $\phi$ with the  
limiting values, $p(\pm 1)=\pm 1$.  
The phase-field model expressed  
in terms of these variables takes the form 
\begin{eqnarray} 
\frac{\partial \phi}{\partial t} &=& 
-\Gamma_\phi\frac{\delta {\cal F}}{\delta \phi}\,+\,\theta(\vec r,t)  
\label{p1}\\ 
\frac{\partial H}{\partial t}&=&\Gamma_H\,\nabla^2 
\frac{\delta {\cal F}}{\delta H}-\vec\nabla\cdot \vec q(\vec r,t) 
\label{p2} 
\end{eqnarray} 
which is a form similar  
to Model C of Halperin, Hohenberg,  
and Ma \cite{Haletal}, i.e. with coupled non-conserved ($\phi$) 
and conserved ($H$) order parameters, 
which is most naturally suited to add fluctuations. 
The fact that $H$ is conserved, 
which follows from Eq. \ref{p2}, 
simply reflects the fact that the total energy  
in a given volume is conserved in the  
absence of energy fluxes through the surfaces 
bounding this volume. 
Next, the free-energy is defined by 
\begin{equation} 
{\cal F}\,=\,\int \,d^3r 
\left[\frac{K}{2}\,|\vec\nabla 
\phi|^2\,+\,h_0\,f(\phi) 
\,+\,e_0\,\frac{u^2}{2}\right] \label{Lyapou}, 
\end{equation} 
where $h_0$ and $K$ are constants with units of energy 
per unit volume and per unit length, respectively, 
and $f(\phi)$ is a double well potential with minima 
at $\phi=\pm 1$. Specific choices for $p(\phi)$ and $f(\phi)$ 
will be given in the next section to carry out the computations. 
Finally, the noises are Gaussianly 
distributed with variances  
\begin{eqnarray} 
<\theta(\vec r,t)\theta(\vec r',t)> 
&=&2\Gamma_\phi k_BT_M\delta(\vec r-\vec r')\delta(t-t')\label{thetacorr}\\ 
<q_m(\vec r,t)q_n(\vec r',t)> 
&=&2\Gamma_Hk_BT_M\,\delta_{mn}\delta(\vec r-\vec r')\delta(t-t')\label{qcorr} 
\end{eqnarray} 

Let us now briefly analyze the equilibrium  
bulk and interface fluctuations 
in this diffuse interface model in order to make contact with 
the sharp-interface model of the previous section. 
As is well-known, the probability, 
$P[\phi,u;t]$, of finding  
the system in a given configuration, 
$\phi(\vec r,t)$ and $u=u(\vec r,t)$, at 
time $t$ is governed by a 
generalized Fokker-Planck equation \cite{Haletal} associated with  
the Langevin equations \ref{p1} and \ref{p2}. For a 
general non-equilibrium situation,  
this Fokker-Planck equation has no 
known analytical solution. In equilibrium, however, it 
has a time-independent stationary solution 
\begin{equation} 
P_{eq}\left[\phi,u\right]= 
\frac{1}{Z}\exp\left(-\frac{\cal F}{k_BT_M}\right),\label{Prob} 
\end{equation} 
which allows us to calculate analytically the 
equilibrium Gaussian fluctuations. Here, 
\begin{equation} 
Z\equiv \int {\cal D}\phi {\cal D}u \,\exp\left(-\frac{\cal F}{k_BT_M}\right) 
\end{equation} 
is the equilibrium partition  
function where ${\cal D}\phi$ and ${\cal D}u$ 
denote functional integration over the fields $\phi$ and $u$, 
respectively. Let us first calculate 
the temperature fluctuations in the bulk 
phases. Since $\phi$ is constant in the solid 
or liquid, only the term $\sim u^2$  
in the integrand of ${\cal F}$ 
needs to be kept. 
Consequently, Eq. \ref{Prob}  
implies that the fluctuation of $u$  
inside a small volume $\Delta V$ is given by 
\begin{eqnarray} 
<u^2>&=&\int_{-\infty}^{+\infty} du\,u^2\, 
\exp\left[-\frac{\Delta Ve_0}{k_BT_M}\,\frac{u^2}{2}\right]/ \\ \nonumber 
& &\int_{-\infty}^{+\infty} du 
\exp\left[-\frac{\Delta Ve_0}{k_BT_M}\,\frac{u^2}{2}\right], 
\end{eqnarray} 
which yields at once the result, 
\begin{equation} 
<u^2>\,=\,\frac{k_BT_M}{e_0\Delta V}\label{ufluc2} 
\end{equation} 
Now comparing Eq. \ref{ufluc2} with 
Eq. \ref{tempfluc} allows us to determine 
\begin{equation} 
e_0=\frac{L^2}{T_Mc}.\label{e0} 
\end{equation} 
This result can be obtained, alternatively, by 
comparing the phase-field equations (\ref{p2}) and (\ref{qcorr}), 
in a region where $\phi$ is constant, with the sharp-interface equations 
(\ref{e1}) and (\ref{e4}), 
which yields, in addition, 
the expression for the diffusion constant 
\begin{equation} 
D=\frac{\Gamma_H}{e_0}. 
\end{equation} 

Next, the equilibrium  
fluctuations of a stationary interface can be calculated 
provided that we restrict our attention 
to wavelengths that are large compared to the  
width of the spatially diffuse 
interface region. 
Let us consider the fluctuations about 
a flat interface in the plane $z=0$. 
For a small amplitude 
deformation, $\zeta(\vec r_\perp)$, which 
varies slowly on the scale of the interface thickness, 
the phase-field can be 
approximated in the form 
\begin{equation} 
\phi(\vec r)\approx \phi_0(z-\zeta(\vec r_\perp)), 
\label{phiapprox} 
\end{equation} 
where $\phi_0(z)$ 
is the solution of the one-dimensional 
stationary interface problem 
\begin{equation} 
K\frac{d^2\phi_0(z)}{dz^2}\,+\,h_0\,f_\phi\left(\phi_0(z)\right)=0, 
\label{1dfront} 
\end{equation} 
where we have defined $f_\phi\equiv df/d\phi$.
We can then evaluate 
the gradient term in Eq. \ref{Lyapou} using 
Eq. \ref{phiapprox}, which yields 
\begin{equation} 
\vec\nabla\phi(\vec r)\approx \frac{d\phi_0}{dz} 
\left[\hat z - \vec \nabla_{\perp}\zeta(\vec r_\perp)\right] 
\label{expand} 
\end{equation} 
Substituting Eq. \ref{expand} into Eq. \ref{Lyapou}, 
we obtain at once that the probability  
distribution of interface 
fluctuations is given by 
\begin{equation} 
P\left[\zeta(\vec r_\perp)\right]=\frac{1}{Z}\exp\left( 
-\frac{\gamma}{k_BT_M}\, 
\int d^2r \frac{1}{2}|\vec \nabla_{\perp}\zeta(\vec r_\perp)|^2 
\right)\label{Pint}, 
\end{equation} 
where 
\begin{equation} 
Z=\int {\cal D}\zeta \exp\left( 
-\frac{\gamma}{k_BT_M}\, 
\int d^2r \frac{1}{2}|\vec \nabla_{\perp}\zeta(\vec r_\perp)|^2 
\right), 
\end{equation} 
and 
\begin{equation} 
\gamma =\sqrt{Kh_0}\, 
\int_{-\infty}^{+\infty} \,d\bar z 
\left[\frac{d\phi_0}{d\bar z}\right]^2\equiv  
\sqrt{Kh_0}\,I 
\label{gammapf} 
\end{equation} 
is the surface energy; the integral $I$  
defined in terms of the dimensionless variable 
$\bar z=z\sqrt{h_0/K}$ is a numerical constant 
that depends on the form of $f(\phi)$. 
The result of Eq. \ref{intfluc} stated earlier 
is now simply obtained by changing  
variables from $\zeta(\vec r_\perp)$  
to $\zeta_k$ in the probability distribution above,  
and by using this distribution to 
calculate $<\zeta_k\zeta_{-k}>$, which 
only involves a Gaussian integral. 
This simple exercise shows that the interface 
fluctuations in the phase-field model are identical
to those of the sharp-interface model on scales 
larger than the interface thickness, i.e. 
$k^{-1}\gg \sqrt{K/h_0}$, as one would 
naively expect. Another way to see this 
correspondence is to note that (\ref{Pint}) is 
identical to the probability distribution of small amplitude 
fluctuations in the sharp-interface model. 
In this model, the probability of an arbitrary 
fluctuation of the interface 
is $\sim \exp(-\gamma \int da/k_BT_M)$  
where $da=d^2r \sqrt{1+|\vec \nabla_\perp\zeta|^2}$ 
is the element of surface area. For a small amplitude 
fluctuation, the square root can be 
expanded to first order which yields at once 
the distribution (\ref{Pint}). 
Finally, Eqs. \ref{e0} and \ref{gammapf} can be used to 
relate the parameters of the phase-field model to the 
capillary length, $d_0=\gamma T_Mc/L^2$,  
which yields  
\begin{equation} 
d_0=I\sqrt{\frac{Kh_0}{e_0^2}} 
\end{equation} 
 
\section{Numerical implementation} 
 
\subsection{Choice of functions and scalings} 
 
To carry out numerical simulations, 
it is convenient to choose the functions 
\begin{eqnarray} 
f(\phi)&=&-\phi^2/2+\phi^4/4\label{fdef},\\ 
p(\phi)&=&15(\phi-2\phi^3/3+\phi^5/5)/8\label{pdef}, 
\end{eqnarray} 
where (\ref{fdef}) is the standard quartic 
form of double well potential and  
the form (\ref{pdef}) 
has the advantage that it preserves the minima  
of $\phi$ at $\pm 1$ independently 
of the local value of $u$ \cite{CagChe}.  
The one-dimensional stationary profile 
solution of Eq. \ref{1dfront} is then given by 
\begin{equation} 
\phi_0(z)=-\tanh\left(\frac{z}{\sqrt{2}W}\right) 
\end{equation} 
where, 
\begin{equation} 
W= \sqrt{\frac{K}{h_0}}, 
\end{equation} 
is the interface thickness. Evaluating 
the integral in Eq. \ref{gammapf} with the above form 
of $\phi_0(z)$ yields $I=2\sqrt{2}/3$. 
 
It is useful to express the phase-field 
equations in a dimensionless form that  
minimizes the number of computational parameters and that renders 
the interpretation of the noise magnitude 
in the phase-field model more transparent.  
For this purpose, it is useful  
to define, in addition to $W$, the time 
\begin{equation} 
\tau= \frac{1}{\Gamma_\phi\,h_0}, 
\end{equation} 
which characterizes the 
relaxation of $\phi$ to one of its local 
minima, and the coupling constant \cite{note} 
\begin{equation} 
\Lambda= \frac{e_0}{Jh_0}=\frac{I}{J}\,\frac{1}{\bar d_0}, 
\end{equation} 
expressed in terms of the scaled 
capillary length 
\begin{equation} 
\bar d_0=\frac{d_0}{W}. 
\end{equation} 
Here, $J=16/15$ is a constant whose value 
is fixed by the choice of $p(\phi)$ \cite{KarRap}. 
We then measure all lengths in units of $W$ and time 
in units of $\tau$, and define 
accordingly new dimensionless  
coordinates, diffusivity,  
and noise variables,  
via the substitutions 
\begin{eqnarray} 
\vec r/W 
&\rightarrow& 
\vec r \label{t1}\\ 
t/\tau 
&\rightarrow& 
t \label{t2}\\ 
D\tau/W^2 
&\rightarrow& 
D\label{t3}\\ 
\tau\theta 
&\rightarrow& 
\theta\label{t4}\\ 
\frac{\tau}{e_0W}\,\vec q 
&\rightarrow& 
\vec q\label{t5} 
\end{eqnarray} 
Transforming the phase-field equations \ref{p1} and 
\ref{p2} with the help of  
these substitutions, and using the fact  
that $\delta(\vec r-\vec r')$ and $\delta(t-t')$ 
on the r.h.s. of equations (\ref{thetacorr}) and (\ref{qcorr}) 
have dimensions of (length)$^{-d}$, where 
$d$ is the dimension, and inverse time, 
respectively, we obtain the dimensionless form 
\begin{eqnarray} 
\frac{\partial \phi}{\partial t} &=& 
\nabla^2 \phi+\phi-\phi^3-\Lambda\,u\,(1-\phi^2)^2 
~+~\theta(\vec r,t) \label{pf1}\\ 
\frac{\partial u}{\partial t}&=& D\,\nabla^2 u\, 
+\,\frac{1}{2}\frac{\partial p(\phi)}{\partial t} 
~-~\nabla\cdot\vec q(\vec r,t)\label{pf2} 
\end{eqnarray} 
with, 
\begin{eqnarray} 
<\theta(\vec r,t)\theta(\vec r',t)> 
&=&2F_\phi\,\delta(\vec r-\vec r')\delta(t-t')\label{etacorr}\\ 
<q_m(\vec r,t)q_n(\vec r',t)> 
&=&2\,D\,F_u\,\delta_{mn} 
\delta(\vec r-\vec r')\delta(t-t')\label{jcorr} 
\end{eqnarray} 
and the definitions, 
\begin{eqnarray} 
F_{ex}&=&\frac{k_BT_M^2c}{L^2d_0^d} \label{Fex}\\ 
F_u&=&\frac{k_BT_M^2c}{L^2W^d}\,=\, 
\bar d_0^d\,F_{ex}\,\label{Fu}\\ 
F_\phi&=&\Lambda J\, F_u\label{Fphi} 
\end{eqnarray} 
The above definitions allow us  
to relate the magnitude of 
the noise which enters into the 
phase-field model, $F_u$, with the magnitude 
of the noise in experiments, $F_{ex}$. 
Comparing the r.h.s of 
Eq. \ref{Fex}, for $d=3$, with the 
r.h.s. of Eq. \ref{ufluc2}, we can readily see that $F_{ex}$ 
is simply equal to the mean-square fluctuation of  
$u$ inside a microscopic volume $d_0^3$, and is 
a fixed quantity for a given material. 
(Note that $F_{ex}$ 
can also be written in the form $k_BT_M/\gamma d_0^2$, 
which is the square of the ratio of two  
microscopic lengths, $\sqrt{k_BT_M/\gamma}$, and $d_0$.) 
The first equality in 
Eq. \ref{Fu} implies that $F_u$  
is the mean-square fluctuation of $u$ inside 
a microscopic volume $W^3$. The second equality
dictates how to choose $F_u$ in a simulation 
for a given system ($F_{ex}$) and a given 
choice of computational parameter, $\bar d_0$. The 
dependence on the latter quantity has a simple  
physical interpretation. 
Namely, if one chooses 
$\bar d_0$ to be small compared to unity, which is the 
main gain in computational efficiency resulting from 
the reformulated asymptotics of Ref. \cite{KarRap}, 
then one must scale down the magnitude 
of the noise in the phase-field model to keep the fluctuation  
strength in a physical volume $d_0^3$ constant. The 
main practical conclusion here is  
that one still has the computational freedom to 
choose the interface thickness if  
one rescales appropriately 
the noise strength. 
 
\subsection{Discretization} 
 
We first discuss  
the numerical implementation of the model 
in two dimensions and then briefly mention its  
straightforward extension to three 
dimensions. 
The phase-field equations (\ref{pf1}) and (\ref{pf2}) 
are discretized on a $N\times N$ 
square lattice of spacing $\Delta x=\Delta z$ 
using centered finite difference 
formulae, as described in Ref. \cite{KarRap}, 
and the equations are time-stepped using a first order 
Euler scheme with a time step $\Delta t$.  
The only new elements here are the noises. 
To see how to discretize them, 
let $i\Delta x$ and $j\Delta z$ denote 
the position on the lattice 
along $x$ and $z$, respectively. 
For the non-conserved noise, we generate 
one random number per lattice site, 
$\theta_{ij}$, chosen from a Gaussian  
distribution with a variance 
\begin{equation} 
<\theta_{ij}\theta_{i'j'}>\,=\,\frac{2F_\phi}{\Delta t\,\Delta x^2}\, 
\delta_{ii'}\delta_{jj'}, 
\label{disctheta} 
\end{equation} 
where the factors $1/\Delta t$ and $1/\Delta x^2$ on 
the r.h.s. of Eq. \ref{disctheta} 
are related to the inverse time and the inverse area 
(inverse volume in three dimensions) scalings of 
$\delta(t-t')$ and $\delta(\vec r-\vec r')$, respectively, 
in the correlation of the noises. 
$\theta_{ij}$ is then added to the deterministic 
part of the r.h.s. of Eq. \ref{pf1} discretized at site $(i,j)$. 
 
To discretize the conserved noise, we define  
by $q_{x,ij}$ the current on the bond that links 
site $(i,j)$ with site $(i+1,j)$, and by $q_{z,ij}$ 
the current on the bond that links site $(i,j)$ 
and $(i,j+1)$. We then generate at each time 
step two independent random numbers per site, 
$q_{x,ij}$ and $q_{z,ij}$, chosen 
from a Gaussian distribution with a variance 
\begin{equation} 
<q_{m,ij}q_{n,i'j'}>\,=\,\frac{2DF_u}{\Delta t\,\Delta x^2}\, 
\delta_{mn}\delta_{ii'}\delta_{jj'}, 
\label{discq} 
\end{equation} 
The divergence of the current on the r.h.s. of 
Eq. \ref{pf2} at site $(i,j)$ is then discretized in the form 
\begin{equation} 
\left(\vec \nabla \cdot \vec q\right)_{ij} 
=\left[q_{x,ij}-q_{x,i-1j}+q_{z,ij}-q_{z,ij-1}\right]/\Delta x 
\label{discdiv} 
\end{equation} 
In three-dimensions, the only changes involve 
replacing $\Delta x^2$ by $\Delta x^3$ in 
the denominator of the r.h.s. of equations (\ref{disctheta}) 
and (\ref{discq}), and to generalize (\ref{discdiv}) to 
\begin{eqnarray} 
& &\left(\vec \nabla \cdot \vec q\right)_{ijk} 
=\left[q_{x,ijk}-q_{x,i-1jk}+q_{y,ijk}-q_{y,ij-1k}\right.\nonumber\\ 
& &~~~~~~~~~~~~~~~~~~~~~~~~~~~~~~~ 
\left.+q_{z,ijk}-q_{z,ijk-1}\right]/\Delta x 
\end{eqnarray} 
 
\subsection{Planar interface fluctuation spectrum} 
 
As a non-trivial  
test of the numerical implementation of the phase-field 
model, we first calculate the fluctuation spectrum 
of a one-dimensional stationary interface 
in thermal equilibrium and compare this 
spectrum to the analytical prediction (\ref{intfluc}). 
With length measured in units of $W$,  
(\ref{intfluc}) becomes  
\begin{equation} 
S(k)=\frac{F_u}{\bar d_0k^2} 
\label{intfluq} 
\end{equation} 
where $\bar d_0=I/(J\Lambda)=5\sqrt{2}/8\Lambda$ 
for the present choice of phase-field model. 
 
To calculate $S(k)$, phase-field simulations 
were carried out with periodic boundary conditions 
in $x$ on a lattice of size $512\times50$ with $\Delta x=0.8$. 
We used initial conditions that correspond  
to a flat interface inside a system uniformly 
at the melting temperature, which corresponds to 
choosing $\phi=\phi_0(z)=-\tanh(z/\sqrt{2})$ and $u=0$. 
The interface profile, $\zeta(x)$, 
is defined by $\phi(x,\zeta(x))=0$, and is calculated 
by finding the $\phi=0$ contour of the phase-field 
by interpolation at each time step. 
The complex amplitude, $\zeta_k$, is then calculated 
by a one-dimensional fast Fourier transform where 
$\zeta_k$ and $\zeta(x)$ are related by 
\begin{equation} 
\zeta(x)=\int\,\frac{dk}{2\pi} 
\,e^{i kx}\,\zeta_k, 
\end{equation} 
Finally, $S(k)=<|\zeta_k|^2>$ 
is calculated by taking a time average  
of $|\zeta_k|^2$. Long simulations with typically 
$10^5$ to $10^6$ time steps were necessary 
to obtain good statistics. These calculations were carried out  
by using (\ref{pf2}) with both $p(\phi)$ defined by (\ref{pdef}) 
and $p(\phi)=\phi$. With the latter choice, the phase-field equations 
are no longer variational (i.e. derivable from a Lyapounov 
functional), but the sharp-interface limit 
remains identical and the interface can be resolved with 
a larger $\Delta x$, as shown previously \cite{KarRap}.  
The spectra for the two choices of $p(\phi)$ were found to be 
virtually indistinguishable such that only the results 
for $p(\phi)=\phi$ are reported here. In the dendritic growth 
simulations presented below, we will restrict our attention 
solely to the case where $p(\phi)=\phi$ is used as the 
source of latent heat in the 
heat equation. 
 
\begin{figure} 
\def\epsfsize#1#2{0.35#1} 
\newbox\boxtmp 
\setbox\boxtmp=\hbox{\epsfbox{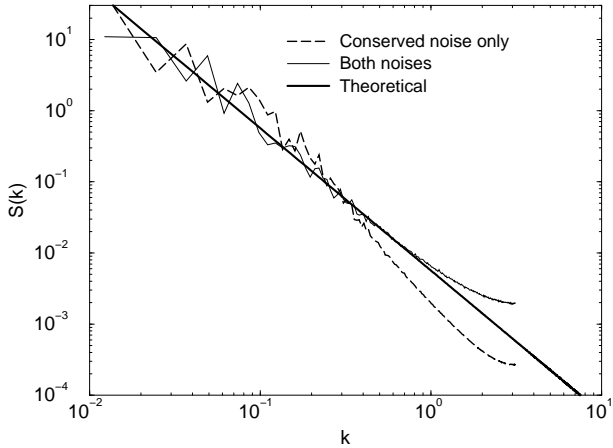}} 
\rotr{\boxtmp} 
\vspace{0.5cm} 
\caption{Simulated
spectra of a one-dimensional
interface in thermal equilibrium with 
both non-conserved and conserved noises (thin solid line)
and only the conserved noise (dashed line),
compared to the theoretical prediction of
Eq. \protect\ref{intfluq} (thick solid line).
Length is measured in units of $W$.
Parameters used in simulations are $\Lambda=1$, $D=1$,
and $F_u=0.005$.
} 
\label{equil} 
\end{figure} 
 
Spectra obtained for a typical set of 
computational parameters are compared in Fig. \ref{equil} with the 
analytical prediction (\ref{intfluq}), represented by a thick solid line, 
both for the case where the 
non-conserved and conserved noises are added to 
the phase-field equations (thin solid line with 
$F_u \ne 0$ and $F_{\phi} \ne 0$) and for  
the case where the non-conserved  
noise is switched off in the $\phi$ equation  
(dashed line with $F_u \ne 0$ but $F_{\phi}=0$). 
With both noises present,  
the calculated spectrum agrees well with  
the theoretical prediction up to a cutoff in $k$ 
of order unity (corresponding to a wavelength comparable 
to the interface thickness in physical units). 
With only the conserved noise present ($F_{\phi}=0$), the
simulated spectrum follows initially well
the predicted $1/k^2$ law with increasing $k$, but
then drops off rapidly to a very small 
amplitude at large $k$.
This drop off is consistent 
with the analytical prediction of 
Ref. \cite{Karma} and is due to
the extra dissipation at the interface that damps 
out short scale fluctuations.
 
\subsection{Incorporation of anisotropy} 
 
In order to investigate dendritic sidebranching 
in the next section, we incorporate anisotropy  
as other authors \cite{Kob:Sim,Wanetal} 
by letting the coefficient of the gradient energy term in the 
free-energy depend on the normal to the solid-liquid 
interface, $\hat n=\vec\nabla\phi/|\vec\nabla\phi|$.
Following this change, (\ref{pf2}) remains 
unchanged and (\ref{pf1}) becomes 
\begin{eqnarray} 
& & \,f_k({\hat n})\,\partial_t \phi= 
\phi-\phi^3-\Lambda\,u\,(1-\phi^2)^2 
\nonumber\\ 
& &+\,{\vec \nabla}\cdot(f_s({\hat n})^2{\vec \nabla} \phi) 
+\sum_{m=x,z} 
\partial_m\left(|\vec \nabla \phi|^2 f_s({\hat n}) 
\frac{ \partial f_s({\hat n})}{\partial 
(\partial_m\phi)}\right)\nonumber\\ 
& &+\,\theta(\vec r,t)\label{pf1b} 
\label{pf1ani} 
\end{eqnarray} 
where we have defined the anisotropy 
function for a crystal with an underlying 
cubic symmetry 
\begin{equation} 
f_s({\hat n})= 
  1-3\epsilon_4 
+ 4 \epsilon_4 (\partial_x^4\phi+\partial_z^4\phi)/ 
|\vec \nabla\phi|^4 \label{ani} 
\end{equation} 
As in our previous study of dendritic growth 
without noise \cite{KarRap}, we use the result of a reformulated 
asymptotic analysis of the phase-field model together 
with a method to compute lattice corrections to the 
surface energy and kinetic anisotropies. Moreover, 
we focus on a choice of computational parameters 
that makes $1/\mu$ vanish in the interface  
condition (\ref{e3}). 
The effective anisotropy of the 
phase-field model, which includes lattice 
corrections, is given at order $\Delta x^2$ by 
\begin{equation} 
\epsilon_e=\epsilon_4-\Delta x^2/240 
\end{equation} 
Here, we use $\Delta x=0.8$,  
and input the value $\epsilon_4=0.03266$ 
into Eq. \ref{ani} to obtain an effective 
3\% anisotropy when comparing our results to 
the sharp-interface solvability theory. 
To make the interface kinetic contribution vanish 
at the same order  
we choose  
\begin{equation} 
f_k(\hat n)= \frac{1-3\delta 
+ 4 \delta (\partial_x^4\phi+\partial_z^4\phi)/ 
|\vec \nabla\phi|^4}{1+\delta}, \label{ani2} 
\end{equation} 
where the value of $\delta$ is computed, 
together with an order $\Delta x^2$ 
correction to $\Lambda$, in order to make $1/\mu$ 
vanish in (\ref{e3}), as described 
in \cite{KarRap}. The resulting 
computational parameters are 
summarized in Table I.  
 
Lastly, in terms of our 
dimensionless units, where length, time, and velocity, 
are scaled in units of $W$, $\tau$, and $W/\tau$, respectively, 
and without interface kinetics, 
the thin-interface limit of the phase-field model 
is the standard free-boundary problem: 
\begin{eqnarray} 
\partial_t u& =& D\, \nabla^2 u 
-\vec \nabla \cdot {\vec q} 
\label{es1}\\ 
V_n & =& -\,D\, \hat n \cdot \left( \vec \nabla u|_l -\vec \nabla u|_s 
\right) 
\,+\,\,\hat n \cdot \left(\vec q|_l - \vec q|_s \right)\label{es2} \\ 
u_I & =& -\bar d_0(1-15\epsilon_e\cos \,4\alpha)\,\kappa 
\label{es3} 
\end{eqnarray} 
where $\vec q$ is the same noise as in the phase-field 
model and $\alpha=\cos^{-1}(\hat z\cdot \hat n)$ is 
the angle of the normal measured from the $z$-axis. 
 
\begin{table} 
\caption{List of the phase-field computational parameters  
used in dendritic growth simulations. These parameters 
yield an effective 3\% anisotropy in surface energy 
and a diverging interface kinetic coefficient $\mu$  
as defined here in Eq. \protect\ref{e3}.} 
\begin{center} 
\begin{tabular}{cc} 
$\Delta x$ & 0.8 \\ 
$\Delta t$ & 0.06 \\ 
$D$ & 2 \\ 
$\Lambda$ & 3.268 \\ 
$\bar d_0$ & 0.27 \\ 
$\epsilon_4$ & 0.03266 \\ 
$\epsilon_e$ & 0.03 \\ 
$\delta$ & 0.046 \\ 
\end{tabular} 
\end{center} 
\label{lattice} 
\end{table} 
 
\section{Dendritic sidebranching} 
 
In this section, we simulate the phase-field model 
defined by Eqs. \ref{pf2} and \ref{pf1b} to investigate
sidebranching characteristics for different 
noise levels and a fixed dimensionless  
undercooling $\Delta\equiv (T_M-T_\infty)/(L/c)=0.55$, 
where $T_\infty$ is the initial temperature 
of the melt. We then compare these results  
quantitatively with the predictions of  
the linear WKB theory that corresponds to the 
sharp-interface model defined  
by Eqs. \ref{es1}-\ref{es3}.  
 
\subsection{Numerical results} 
 
Test simulations were first carried out 
with both noises, $\theta$ and $\vec q$, and with 
only the conserved noise $\vec q$. We found that 
time-averaged sidebranching characteristics were identical 
for the two cases within our numerical resolution. 
This finding shows that fluctuations  
which become amplified to produce sidebranches 
are on lengthscales much larger 
than the interface thickness, and thus driven solely 
by the bulk noise in agreement with 
expectation (see section I and \cite{Karma}). 
Consequently, all the  
results presented in this section were 
obtained with simulations where noise 
is added only to the heat transport equation (\ref{pf2}).
This represents a non-negligible computational saving 
for long simulation runs (i.e. 2 instead of 3 random 
numbers per site at each time step). 
 
\begin{figure} 
\def\epsfsize#1#2{0.35#1} 
\newbox\boxtmp 
\setbox\boxtmp=\hbox{\epsfbox{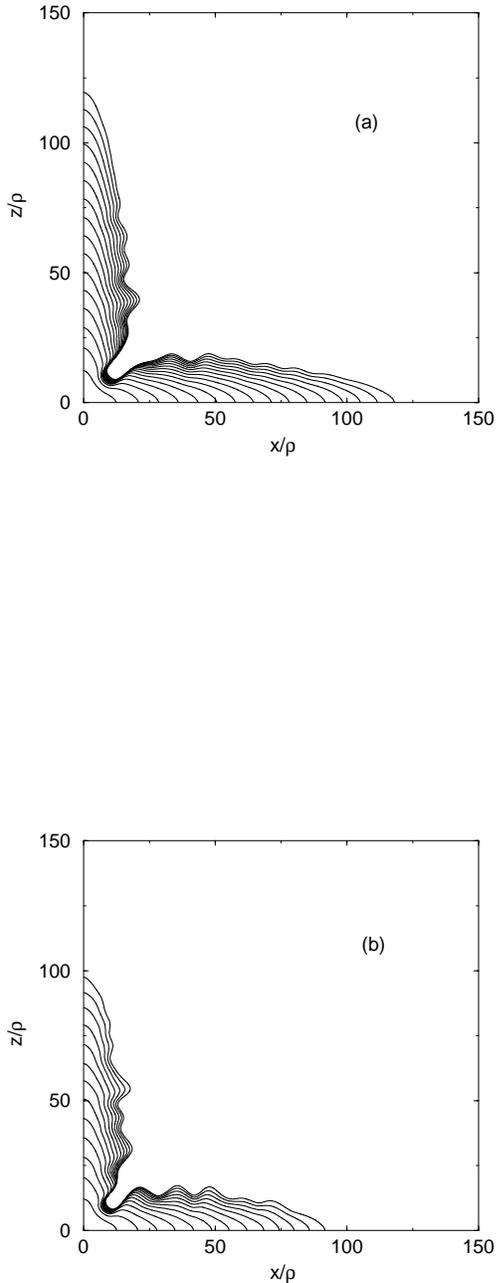}} 
\rotr{\boxtmp} 
\vspace{0.5cm} 
\caption{Morphological development of a solid seed for  
$\Delta=0.55$ and two different conserved 
noise amplitudes: (a) $F_u=10^{-4}$, and (b) $F_u=10^{-3}$. 
Other computational parameters are listed in Table 1. 
The interfaces are plotted every 16000 iterations. 
} 
\label{dynlow} 
\end{figure}

The development of a dendrite and its
sidebranches from a small initial seed is illustrated 
in Figs. \ref{dynlow}a and \ref{dynlow}b  
for two different noise levels.  
These particular simulations were 
carried out on a large $1200\times 1200$ lattice 
with, as initial condition, $u=0$ and $\phi=1$ inside a 
small circle in the lower left hand corner of the quadrant  
and $u=-\Delta$ and $\phi=-1$ outside this circle. 
Note that in Fig. \ref{dynlow}b the noise is 
sufficiently large to disturb the 
steady-state growth of the tip, which can be deduced 
from the fact that the vertical branch has  
outgrown the horizontal branch in this case. 
Since the tips do not interact via the diffusion field 
at this undercooling, i.e. 
the separation between the tips is much larger than 
the diffusion length, this difference can only be due 
to noise. This effect is negligible 
for the smaller noise level (Fig. \ref{dynlow}a) where 
the two tips grew at nearly the same rate.  
 
\begin{figure} 
\def\epsfsize#1#2{0.35#1} 
\newbox\boxtmp 
\setbox\boxtmp=\hbox{\epsfbox{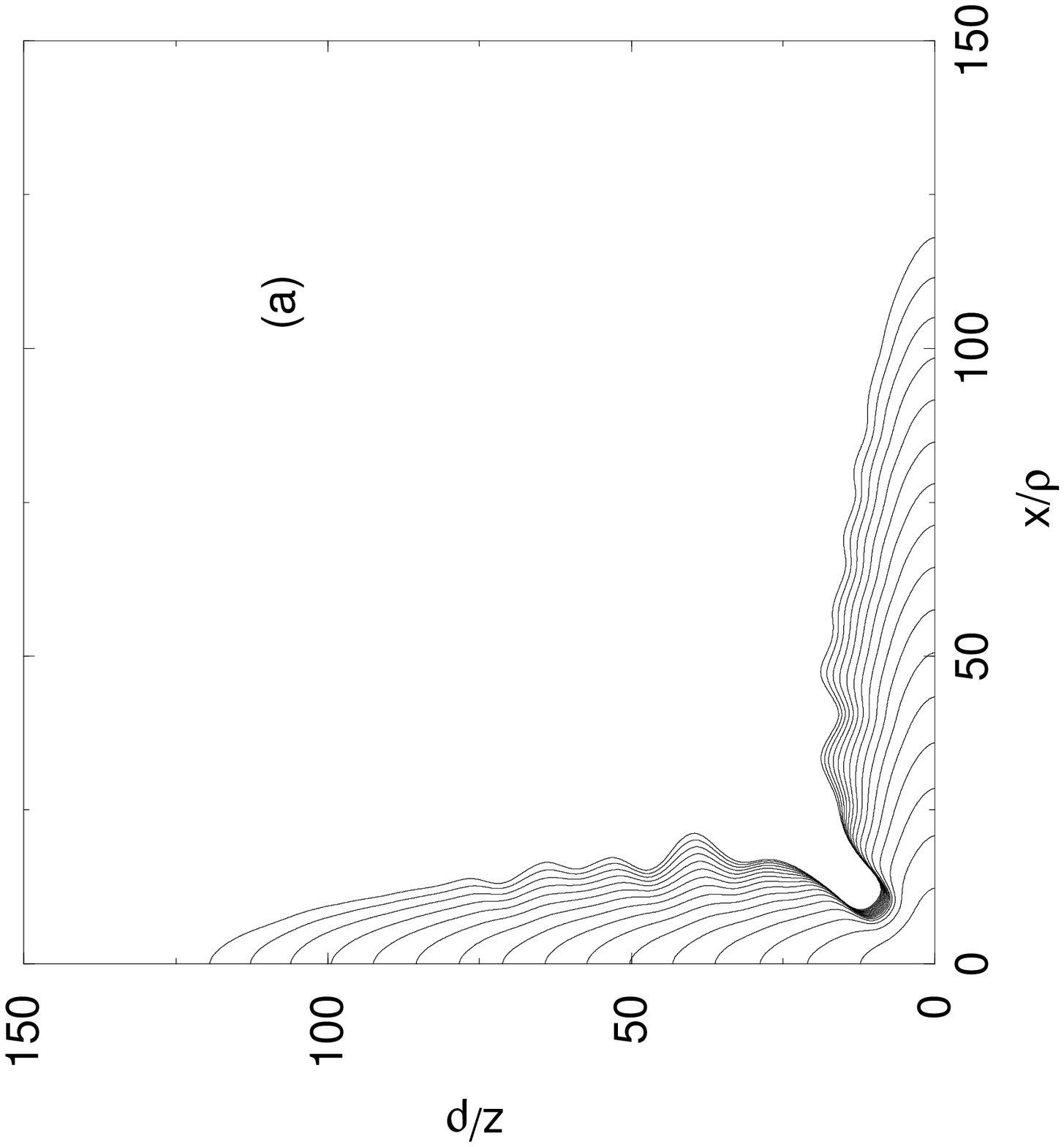}} 
\rotr{\boxtmp} 
\vspace{0.5cm} 
\end{figure} 
 
To investigate sidebranching, we 
restrict our attention to small noise levels  
($F_u=2.5 \times 10^{-5}$ and $F_u=2.5 \times 10^{-4}$) 
with a well-defined steady-state tip structure. 
The symmetric growth of one tip about the $z$-axis 
(i.e. half the dendrite with reflection symmetry) 
is simulated on lattices of size $300 \times 400$ and 
$300 \times 600$ for, respectively, the larger 
and smaller noise amplitude (where sidebranches form 
further behind the tip).  
As in Ref. \cite{KarRap}, we  
periodically translate the entire structure 
in the opposite direction of growth to allow 
long simulation runs to be carried out  
in the smallest lattice size possible. 
Of course, we make sure that the sidebranching 
activity is not affected by this  
procedure by choosing a reasonable 
buffer larger than the diffusion length, and by carrying out 
test runs with larger lattice sizes. 
The constraint of symmetric growth prevents us from 
investigating the correlation of the sidebranching 
activity on opposite sides of the growth axis, which has 
been examined experimentally. However, it permits 
a more efficient investigation of the  
sidebranching amplitude and wavelength 
which can be compared to 
analytical predictions. 
 
\begin{figure} 
\def\epsfsize#1#2{0.35#1} 
\newbox\boxtmp 
\setbox\boxtmp=\hbox{\epsfbox{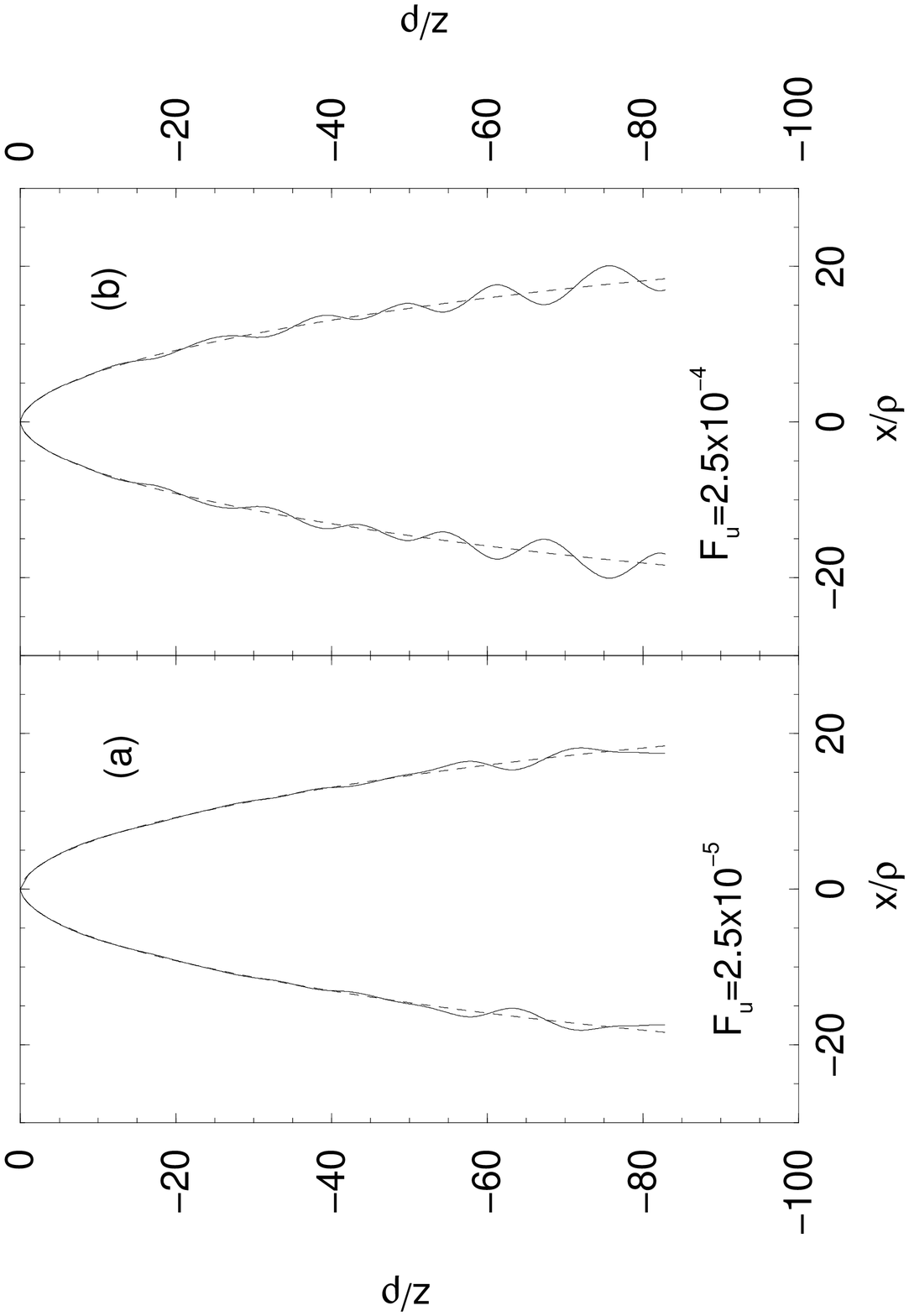}} 
\rotr{\boxtmp} 
\vspace{0.5cm} 
\caption{Snapshots of the time-dependent 
dendrite shapes (solid lines) in long simulation 
runs that focus on the growth of one tip for $\Delta=0.55$ and 
the parameters of Table I. The noiseless 
shape (dashed line) is superimposed for comparison. 
The noise levels are $F_u=2.5 \times 10^{-5}$ in (a) and 
$F_u=2.5 \times 10^{-4}$ in (b). Note that  
sidebranches form further behind the 
tip for the smaller noise level. 
} 
\label{typshap} 
\end{figure}

\begin{figure} 
\def\epsfsize#1#2{0.35#1} 
\newbox\boxtmp 
\setbox\boxtmp=\hbox{\epsfbox{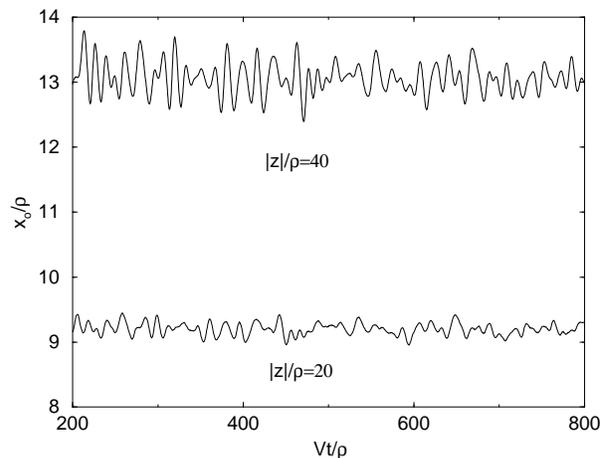}} 
\rotr{\boxtmp} 
\vspace{0.5cm} 
\caption{Horizontal position of the interface 
measured from the vertical growth axis as  
a function of dimensionless time at 20 and at 
40 tip radii behind the tip. 
The parameters are the same  
as in Fig. \protect\ref{typshap}b. 
} 
\label{timetrace} 
\end{figure}

\begin{figure} 
\def\epsfsize#1#2{0.35#1} 
\newbox\boxtmp 
\setbox\boxtmp=\hbox{\epsfbox{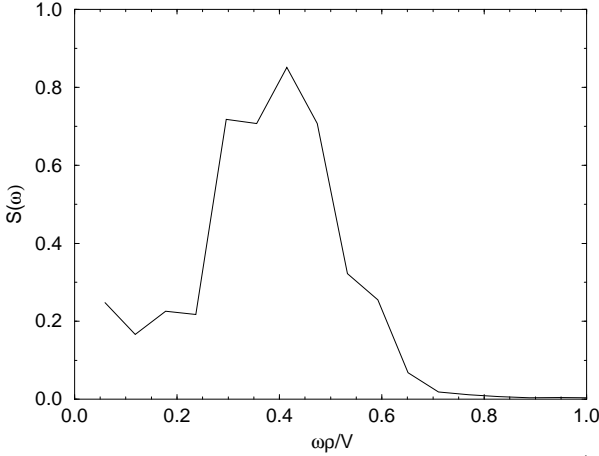}} 
\rotr{\boxtmp} 
\vspace{0.5cm} 
\caption{Noise averaged powerspectrum  
of $x(z,t)$ 
at $|z|/\rho=40$ for $F_u=2.5 \times 10^{-5}$. 
} 
\label{spectrum} 
\end{figure}

To calculate these two quantities, 
we proceed in two steps.  
Firstly, we carry out a simulation without noise 
to obtain a `reference' steady-state (needle crystal) 
shape, without sidebranching.  
It is useful to measure 
this shape by the horizontal distance, $x_0(z)$, of  
the interface measured from the vertical growth axis 
as a function of the distance, $z$, behind the tip.
This distance is calculated by numerical interpolation of 
the $\phi=0$ contour. 
The steady-state operating state of the dendrite 
is defined in terms of the dimensionless  
tip velocity and radius 
\begin{eqnarray} 
\tilde{V}&=&V\bar d_0/D\\ 
\tilde{\rho}&=&\rho/\bar d_0 
\end{eqnarray} 
For the present choice of undercooling and 
anisotropy, we find that $\tilde{V}=V d_0/D\approx 0.011$ 
and $\tilde{\rho}=\rho/d_0\approx21.8$, 
in good agreement with 
the predictions of solvability theory 
\cite{KarRap,solv,KesLev:2D,Mei:2,Ben:2D}. 
Secondly, we add the conserved noise to the heat equation 
and calculate the time-dependent shape, $x(z,t)$, with 
sidebranching present. Snapshots of noisy shapes  
superimposed on the noiseless shape are illustrated 
in Fig. \ref{typshap}. Examples of time traces of 
$x(z,t)$ for two different distances behind the tip 
are shown in Fig. \ref{timetrace}. In addition,  
an example of the noise averaged power  
spectrum of a long time trace 
is shown in Fig. \ref{spectrum}. 
This spectrum was calculated by  
subdividing the complete time interval  
into several equal subintervals, then calculating 
the power spectrum for each subinterval, 
and finally taking the average of  
these power spectra. 
 
\begin{figure} 
\def\epsfsize#1#2{0.35#1} 
\newbox\boxtmp 
\setbox\boxtmp=\hbox{\epsfbox{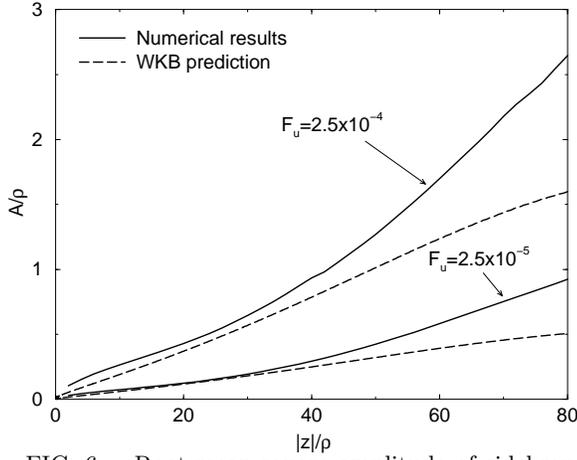}} 
\rotr{\boxtmp} 
\vspace{0.5cm} 
\caption{ 
Root-mean-square  
amplitude of sidebranches as a function of 
distance behind the tip for two different  
noise levels in the simulations (thick solid 
and dash lines). Superimposed are thin solid lines 
corresponding to the analytical predictions  
of Eq. \protect{\ref{Amp}}.} 
\label{growth} 
\end{figure}

\begin{figure} 
\def\epsfsize#1#2{0.35#1} 
\newbox\boxtmp 
\setbox\boxtmp=\hbox{\epsfbox{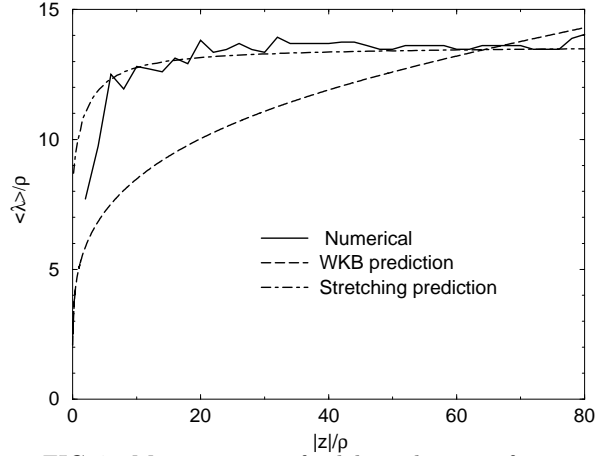}} 
\rotr{\boxtmp} 
\vspace{0.5cm} 
\caption{Mean spacing of sidebranches 
as a function of distance behind the tip 
in the simulation for $F_u=2.5 \times 10^{-5}$ (solid line), 
analytically predicted by Eq. \protect\ref{thspac} 
(dashed line) and predicted on the basis of stretching 
(dotted line). 
} 
\label{spac} 
\end{figure}

In terms of the above quantities, the 
root-mean-square amplitude of sidebranches is 
simply given by: 
\begin{equation} 
A(z)=\sqrt{<(x(z,t)-x_0(z))^2>} 
\end{equation}  
where the average is over time. 
This quantity is plotted vs $z$ in Fig. \ref{growth} 
for two different noise levels. 
To obtain good statistics, we typically simulated  
a total time of $2000 V/\rho$ which 
took 200-350 CPU hours on a high end workstation. 
The mean spacing between sidebranches (sidebranching 
wavelength), $<\lambda(z)>$, can be 
calculated in two ways. One way, which corresponds 
more directly to the way in which this quantity is 
calculated in the WKB theory discussed below, is to define 
\begin{equation} 
<\lambda(z)>=\frac{2\pi V}{\omega_c(z)}\label{way1} 
\end{equation} 
where $\omega_c$  
is the peak frequency 
of the power spectrum of $x(z,t)$, averaged 
over sufficiently long time. An alternate, 
and simpler way, which avoids to 
calculate the power spectrum, is 
to count the number, $N(z)$, of extrema of $x(z,t)$ in 
a long time interval $t_1\le t\le t_2$. 
Simple node counting then leads to the 
relation 
\begin{equation} 
<\lambda(z)>=\frac{2 V(t_2-t_1)}{N(z)}\label{way2} 
\end{equation} 
We have checked  
that these two methods yield nearly identical values 
for the spectrum of Fig. \ref{spectrum}. 
Therefore, we have used Eq. \ref{way2} to calculate 
$<\lambda(z)>$ vs $z$ and the 
result for the lowest noise level 
is shown in Fig. \ref{spac}. 
 
\subsection{Comparison with linear WKB theory} 
 
Langer \cite{Lan}, and Brener and Temkin \cite{Bre:noi}, have 
analyzed noise-induced 
sidebranching in three dimensions 
for specific needle crystal shapes (i.e. $x\sim z^{1/2}$ and 
$x\sim z^{3/5}$). It is straightforward to extend  
their analyses, based on a WKB approach, to an arbitrary needle  
crystal shape, $x_0(z)$, in $d$-dimension \cite{Karmaunp}.
We shall only state here the final 
results necessary to interpret our simulations. 
The expressions for 
the sidebranching amplitude and 
wavelength are given, respectively, by \cite{Karmaunp}:
\begin{eqnarray} 
\bar A(\bar z)&=&\bar S\, 
\exp\left(\frac{2}{3}\left[ 
\frac{\bar x_0^3} 
{3\,\sigma^* \bar z}\right]^{1/2}\right) 
\label{Amp}\\ 
<\bar \lambda(\bar z)>&=&\frac{2\pi}{\bar \omega_c}\,=\,\pi  
\left[\frac{12 \sigma^* \bar z}{\bar x_0}\right]^{1/2} 
\label{thspac} 
\end{eqnarray} 
where we have defined the scaled quantities
$\bar x_0=x_0/\rho$, $\bar z=-z/\rho$, 
$\bar A=A/\rho$, $\bar\lambda=\lambda/\rho$,
$\bar \omega=\omega \rho/V$, 
the dimensionless noise 
amplitude, $\bar S$, given by 
\begin{equation} 
\bar S^2~=~\frac{2F_u\,D}{\rho^{1+d}\,V} 
\,=\,\frac{2F_u}{\bar d_0^d\,\tilde\rho^{1+d}\,\tilde V} 
~~~~~~(d=2,3)\label{bars} 
\end{equation} 
and 
\begin{equation} 
\sigma^*~\equiv~\frac{2D\bar d_0}{\rho^2V}\,=\, 
\frac{2}{\tilde \rho^2\tilde V} 
\label{sigm} 
\end{equation} 
It is easy to check that Eqs. \ref{Amp} 
and \ref{thspac} reduce to 
the earlier results of Refs. \cite{Lan,Bre:noi} if
specific shapes (parabola and 3/5 law) are substituted into them.
Note that if we convert back to dimensional units by 
letting $\rho\rightarrow \rho/W$, $V\rightarrow V\tau/W$,  
and $D\rightarrow D\tau/W^2$,  
in Eq. \ref{bars}, we obtain the expression
\begin{equation} 
\bar S^2~=~ 
\frac{2k_BT_M^2\,c\,D}{L^2\rho^{1+d}\,V}\label{Anoise} 
\end{equation} 
which is dimensionless if one interprets
$L$ and $cT_M$ to have dimension of energy/(length)$^d$. 
Of course, this interpretation is only 
physically meaningful in three dimensions
where Eq. \ref{Anoise} becomes identical to the definition
of $\bar S$ in Ref. \cite{Lan}.
Therefore, in the present study we evaluate 
$\bar S$ directly from Eq. \ref{bars} to 
compare simulations and theory. 
For $\bar d_0=0.27$ (table I) and 
the aforementioned  
selected values, $\tilde V\approx 0.011$  
and $\tilde \rho\approx 21.8$,  
we obtain $\bar S\approx 0.24\,F_u$  
and $\sigma^*\approx 0.383$.  
(Note that $\sigma^*$ is larger here 
than in experiment due to both, the large value of 
anisotropy which produces a pointy tip, and the 
fact that $\sigma^*$ is larger in 2-d than 3-d 
for the same anisotropy.) 
The analytical predictions for the  
sidebranching amplitude 
and wavelength are then simply obtained by 
inputing these values into Eqs. \ref{Amp} and \ref{thspac} 
together with the steady-state interface shape, 
$\bar x_0(\bar z)$, measured in the noiseless phase-field 
simulation (dashed lines in Fig. \ref{typshap}).  
 
Figs. \ref{growth} and \ref{spac} show that  
the amplitude and wavelength measured in the 
phase-field simulations with noise are in 
good overall quantitative agreement with the
analytical predictions even though $\sigma^*$ 
is not much smaller than one. 
The amplitude in the simulations is relatively 
well predicted by Eq. \ref{Amp}, up to a certain 
distance behind the tip after which the 
two curves depart from each other. This departure  
may be due to nonlinear effects which  
become important when $\bar A$ becomes  
of order unity. In addition, it should be emphasized 
that the prefactor of Eq. \ref{Amp} is only known 
up to some multiplicative factor of order unity. 
Consequently, what is more relevant here is that 
the amplitudes in simulation and theory 
are of comparable magnitude, 
rather than the fact that the numerical and theoretical curves in 
Fig. \ref{growth} seem to almost perfectly overlap 
up to $\bar z\simeq 20$, which may be coincidental. 
 
The wavelength in the simulations is only about 30\%  
larger than predicted by Eq. \ref{thspac} 
in the region (20 to 40 tip radii 
behind the tip) where sidebranches becomes visible.
However this wavelength increases initially faster with 
distance behind the tip than predicted.
One possible explanation for this faster rate of  
increase is that perturbations generally get stretched
as they travel along the sides of curved fronts
\cite{Pel,PelCla,Zeletal}.
To test this possibility,
let us calculate the 
purely deterministic change of wavelength of
a perturbation initially at the tip
due to stretching. The
rate of stretching is given by
\cite{Pel,PelCla,Zeletal}:
\begin{equation} 
\frac{1}{\lambda}\frac{d\lambda}{dt} 
=\frac{\partial V_t}{\partial s}\label{stretch} 
\end{equation} 
where $V_t=V\sin\alpha$ is the tangential 
velocity of advection of the perturbation and  
$s$ measures the arclength along the 
interface. Eq. \ref{stretch} is strictly 
valid in the WKB limit where $\lambda$ is 
small compared to the local radius of 
curvature ($1/\kappa$) of the interface. 
We can solve Eq. \ref{stretch} by using 
the change of variable $dt=dz/V$. Eq. \ref{stretch} 
becomes then, $d(\rm{ln}\lambda)=\sin 2\alpha d\alpha/2$, which  
can be easily integrated. Furthermore, using the 
geometrical relation, 
$\cos\alpha=1/[1+(d\bar x_0/d\bar z)^2]^{1/2}$, 
we obtain  
\begin{equation} 
\bar\lambda(\bar z)\,=\,\bar\lambda_{\infty} 
\exp\left[-\frac{1}{2}\,\frac{(d\bar x_0/d\bar z)^2}{1+ 
(d\bar x_0/d\bar z)^2}\right]\label{stonly} 
\end{equation} 
where we have defined $\bar \lambda_{\infty}$ to 
be the saturation value of $\bar \lambda$ 
far behind the tip (i.e $\bar z \rightarrow \infty$). 
In order to test if the 
disagreement between simulation and theory 
in Fig. \ref{spac} is due to stretching near the tip, 
we have plotted in the same figure  
the prediction of Eq. \ref{stonly}  
with $\bar \lambda_{\infty}/\rho=13.5$ fitted from the  
simulation data. We can see that this prediction
indeed improves the agreement with simulation in this region.
This stretching effect is not contained in  
Eq. \ref{thspac} which is only 
valid in the far tip region ($\bar z\gg 1$).
The slower increase in $<\bar \lambda>$ 
predicted by Eq. \ref{thspac} 
is due to the distinct effect that 
lower frequency perturbations survive at a larger 
distance from the tip \cite{BBL,Lan}. It should be
in principle possible to carry out a more elaborate
WKB analysis with noise that includes both the latter effect and
stretching. Such an analysis will yield the same
prediction as Eq. \ref{thspac} in the far tip region,
and (presumably) an improved wavelength prediction 
closer to the tip. 

Finally, we note that the
sidebranching wavelength is about an order of
magnitude larger than the tip radius in our
simulations, whereas it is only a factor
of two or three in the classic experiments of Huang and Glicksman 
in sunccininitrile \cite{HG}. 
This difference is due to the fact that $\sigma^*$
is much larger here than in these experiments
because of the larger value 
of anisotropy used in
simulations.

\section{Conclusions} 
 
We have presented 
a phase-field model of the solidification of 
a pure melt that incorporates  
thermal noise quantitatively. 
From a computational standpoint, there 
are two main conclusions regarding the 
incorporation of this noise. 
Firstly, one can retain 
the freedom to choose the interface 
thickness at will as long as the noise magnitude ($F_u$)
that enters in the phase-field model is 
scaled appropriately (Eq. \ref{Fu}).
Therefore, it remains possible
to carry out dramatically more efficient computations
without interface kinetics 
by choosing $\bar d_0$ substantially smaller than unity,
as in our earlier studies without noise \cite{KarRap}. 
Secondly, for typical 
growth conditions at low undercooling (and, more generally, 
below a critical velocity that depends on the  
attachment kinetic coefficient $\mu$), 
the conserved noise in the heat current  
is the most relevant one. This noise drives 
long-wavelength interface fluctuations that become 
amplified to a macroscopic scale by the morphological 
instability on the sides of dendrites. In contrast, 
the non-conserved noise in the evolution 
equation for $\phi$ drives short-wavelength 
fluctuations that are damped and do not 
affect sidebranching, as predicted by a sharp-interface
analysis \cite{Karma} and confirmed by  
our simulations. Consequently, this noise can 
be left out in computations below this 
critical velocity. 
 
We have applied this model 
to carry out a detailed quantitative study of  
the initial stage of sidebranch formation
during dendritic growth.
The main conclusion is that the 
sidebranching characteristics (root-mean-square 
amplitude and the mean sidebranch 
spacing) are reasonably well predicted by the 
existing linear WKB theory of  
noise amplification, even though the value of
$\sigma^*$ in our simulations is 
not small. A more stringent test of this theory 
would therefore require to extend 
the present study to a range of smaller anisotropy,
and hence smaller $\sigma^*$,
where a comparison between WKB theory
and simulations is more justified.
Finally, since this study
has been restricted to two dimensions, we cannot yet 
answer the important question of whether thermal 
noise alone is responsible for the 
sidebranching activity observed  
in experiment. Simulations
in three dimensions should provide a
clear cut answer to this question. 
 
\begin{center} 
{\bf ACKNOWLEDGMENTS} 
\end{center} 
We thank Mathis Plapp for valuable suggestions concerning
the numerical treatment of the conserved noise.
This work was supported by DOE grant No DE-FG02-92ER45471 
and benefited from CRAY T3E time at the National Energy 
Resources Supercomputer Center,
and time allocation at the Northeastern 
University High-Performance Computing Center (NU-HPCC).


\begin{thebibliography}{999}
\bibitem{note} The coupling constant is denoted 
by $\Lambda$, in place of $\lambda$  
in Ref. \cite{KarRap}, since $\lambda$ is used here to 
denote the wavelength of fluctuations or 
the spacing between sidebranches.
\end{thebibliography}
\end{document}